\documentclass[aps,floats,amssymb,twocolumn,prb]{revtex4}
\usepackage{calc}
\usepackage{psfrag}
\usepackage{graphicx}
\usepackage{color}

\begin{document}

\title{Meron deconfinement in the quantum Hall bilayer at intermediate distances}

\author{M. V. Milovanovi\'c$^1$}
\author{E. Dobard\v{z}i\'c$^2$}
\author{Z. Papi\'c$^3$}

\affiliation{$^1$ Scientific Computing Laboratory, Institute of Physics Belgrade, University of Belgrade, 11080 Belgrade, Serbia}
\affiliation{$^2$ Faculty of Physics, University of Belgrade, 11001 Belgrade, Serbia}
\affiliation{$^3$ School of Physics and Astronomy, University of Leeds, Leeds, LS2 9JT, United Kingdom}

\begin{abstract}
Quantum Hall bilayer phase diagram with respect to interlayer distance bears a remarkable similarity with phase diagrams of strongly correlated systems as a function of doping, with magnetic ordering on the one end and Fermi-liquid-like behavior on the other. Moreover, it has been suggested~\cite{msr} that a BCS correlated state of composite fermions with $p$-wave pairing may exist in the intermediate region. 
In the same region, an exact diagonalization study in the torus geometry \cite{park} 
pointed out the existence of state(s) with pseudospin spiraling order. 
Here we reconcile these two descriptions of the intermediate state by considering the underlying bosonic representation of the composite fermion paired state in the long distance limit, and by performing extensive exact diagonalizations on the torus.
We argue that the spiraling states belong to the manifold of degenerate ground state(s), and are a consequence of Bose 
condensation of the quasiparticles (with critical algebraic correlations) at non-zero momenta in the
two pseudospin states. The spiraling states, generated in this way as spin-textures, can be identified with meron-antimeron
constructions. Thus, merons -- the fractionally charged vortex excitations of the XY magnetically ordered state -- constitute
some of the topological sectors. It follows that merons are deconfined in the intermediate state, and allow for a smooth transition
between the magnetically ordered and Fermi-liquid-like phases, in which they are bound in pairs.

\end{abstract}

\maketitle

\section{Introduction}
Multicomponent fractional quantum Hall (FQH) systems feature rich physics of strongly interacting electrons. In contrast to strong electronic correlations in various (material dependent) crystal settings, in FQH systems the kinetic energy is effectively removed from the problem due to the Landau quantization in strong magnetic fields. This makes the multicomponent FQH systems a simpler (better controlled) setting that is amenable to the investigations of fundamental physics of strongly correlated electrons.

One of the most prominent examples of multicomponent FQH systems is the quantum Hall bilayer (QHB) at total filling factor one. The filling factor $\nu=2\pi\ell_B^2 \rho$ is defined as a product of electron density $\rho$ and the magnetic area $2\pi\ell_B^2$, where $\ell_B=\sqrt{h/eB}$ is the emergent length scale in the magnetic field (``magnetic length")~\cite{prange}. Alternatively, the filling factor represents the ratio of the number of electrons $N$ to the number of magnetic flux quanta $N_\Phi$ through the system. In the case of the $\nu=1$ QHB studied here, there are two quantum wells (``layers") that we label by $\uparrow, \downarrow$, and $N=N_\uparrow + N_\downarrow=N_\Phi$. Note that throughout this work we assume the ordinary electron spin to be fully polarized by the external magnetic field.

Because of the two-fold layer degree of freedom, in general we need to distinguish the interaction between electrons in the same layer (``intra-layer") versus electrons in opposite layers (``inter-layer"). The difference between these two interaction strengths can be tuned experimentally via the parameter called the bilayer distance $d$, which makes the physics in this system much richer compared to an ordinary system of electrons with spin. QHB system has indeed attracted much attention since the time of the theoretical prediction~\cite{fer,moon} and experimental detection~\cite{spie} of the ordered state of excitons (bosons) that forms when $d\lesssim \ell_B$, despite the fact that the fundamental charge carriers in the system are fermions. Perhaps equally remarkable and interesting is the physics at larger (intermediate) distances, typically $1 \lesssim d/\ell_B \lesssim 2$. There, an intermediate phase of paired composite fermions, introduced in Ref. \onlinecite{msr} [see Eqs. (\ref{fpairing1}) and (\ref{sdispairing}) below], is believed to exist. On the other hand, an exact diagonalization study of the system in the torus geometry \cite{park} also pointed out to the existence of pseudospin state(s) with spiraling order in the intermediate region. Here we reconcile these two descriptions of the intermediate region by considering the underlying bosonic representation of the composite fermion paired state in the long distance limit, and by performing extensive exact diagonalization studies on the torus. 

This paper is organized as follows. In Section~\ref{sec:review} we start with an overview of the QHB physics from the point of view of model wave functions. We first discuss the limits of small and large bilayer distances which are well-understood and correspond to the exciton superfluid and the product of two decoupled (composite) Fermi liquid states, respectively. We also review the theoretical evidence for a paired phase in the regime of intermediate bilayer distances, and briefly motivate and summarize our contributions. In Section \ref{sec:picture} we provide a simple physical picture for the formation of the putative intermediate state. Section \ref{sec:numerics} contains results of exact diagonalization studies of the QHB on the torus. In Section \ref{sec:boson} we discuss the torus results from the composite boson viewpoint, and give arguments for meron deconfinement in the intermediate region. We conclude in Section \ref{sec:conc} with a short summary and discussion. Appendices contain further resullts on the intermediate phase.

\section{Bilayer phases and their model wave functions}\label{sec:review}

The QHB at the total filling factor one consists of two quantum Hall mono-layers separated by distance $d$ in the $z$-direction (perpendicular to both layers). The tunneling between the layers can be nearly completely suppressed in experiment. Therefore, we can assume that $N_\uparrow=N_\downarrow$ and each layer is at filling one half. In the following we first give the basic physical picture of the ground state in the limit of either small ($d \lesssim \ell_B$) or large distances $d\gg \ell_B$, followed by the discussion of the intermediate regime $1\lesssim d/\ell_B \lesssim 2$.

\subsection{Small bilayer distances: exciton superfluid}

For small distances, the intra- and inter-layer interactions are of similar strength, and favor each electron in one layer to be directly on top of a hole in the other layer, which drives exciton formation. In this simplified picture we can differentiate between up and down excitons, but due to the quantum-mechanical uncertainty in which layer an electron is, we must include the pseudospin (layer) degree of freedom. The appropriate model function at total filling factor one then reads~\cite{fer,moon}
\begin{equation}
\Psi_{exc}(z_1,\ldots, z_N) = \prod_{i<j}^N (z_i - z_j) e^{-\frac{1}{4\ell_B^2}\sum_k |z_k|^2} \times | \rightarrow \cdots \rightarrow \rangle.
\label{ferro}
\end{equation}
The spatial part of the wave function is written in terms of 2D electron coodinates $z_j = x_j + i y_j$ (regardless of the layer index), and assumes the circular gauge for the magnetic field in the infinite plane. The spatial correlations are encoded via the Laughlin-Jastrow factor $(z_i-z_j)$, and the wave function vanishes at large distances because of the Gaussian term that originates from the single-electron orbitals~\cite{prange}.

Note that the wave function also contains the pseudospin part $| \rightarrow \rightarrow \cdots \rightarrow \rangle$. The pseudospin of each electron points to the same direction, due to the combined effect of the Pauli principle and the Coulomb interaction. The state of each particle $j$ is given by the pseudospin
\begin{equation}
|\rightarrow_j\rangle \equiv \frac{1}{\sqrt{2}} (|\uparrow_j\rangle + e^{i\phi} |\downarrow_j\rangle),
\end{equation}
i.e., the pseudospin is polarized in the $XY$ plane.
This happens because of the capacitor effect which enforces the equality of total charges in the ``up" and ``down" layer, and breaks the $SU(2)$ spin symmetry, assumed to exist at $d=0$, down to $U(1)$. This correlation - condensation was detected in several experiments, e.g., Spielman \emph{et al.} \cite{spie}, although the ``zero-bias peak" anomaly was never as high and sharp as in the similar experiments on superconductors. More recent experiments using Coulomb drag and counterflow measurements are also completely in favor of exciton formation [for a recent review, see Ref. \onlinecite{eis}].

\subsection{Large bilayer distances: product state of two composite Fermi liquids}

When the layers are far apart, it is well-known that each half-filled layer is in the state of the ``Fermi-liquid-like" state \cite{hlr}. For example, the state of the upper layer is given by
\begin{eqnarray}
\nonumber \Psi_{FL} (z_1, \ldots, z_{N_\uparrow}) & = & {\rm det} (e^{\frac{i}{2} (k_m \bar z_n + \bar k_m z_n)}) \prod_{i<j}^{N_\uparrow} (z_i - z_j)^2 \\
&& \times |\uparrow \ldots \uparrow\rangle,
\end{eqnarray}
where we dropped the single-particle Gaussian factors.
The Fermi liquid correlations result from the factor ${\rm det} (e^{i \mathbf{k}_m \mathbf{r}_n})$ that represents the Slater determinant of plane waves. Here, $\{\mathbf{k}\}$ is the set of 2D momenta that define the 2D Fermi sphere and we have substituted $\mathbf{k} \mathbf{r} = (k \bar z+ \bar k z)/2$. The total wave function of the system in this case is just a direct product  of independent layers,
$$
\Psi_{FL} (z_1^\uparrow, \ldots, z_{N_\uparrow}^\uparrow) \otimes \Psi_{FL} (z_1^\downarrow, \ldots, z_{N_\downarrow}^\downarrow).
$$

Notice that both $\Psi_{exc}$ and $\Psi_{FL}$ contain a part which is the product of the differences of $z$'s (the Laughlin-Jastrow part). This part of the wave function fixes the filling factor of the state, which can be seen as follows. On the one hand, the highest exponent in any of the $z$'s is equal to the number of flux quanta through the system, which is the degeneracy of the lowest Landau level (LLL). On the other hand, this number is  equal to the number of electrons in the case of $\Psi_{exc}$, and twice larger than the number of electrons in the case of $\Psi_{FL}$, as follows from the filling factors. Thus, formally speaking the Laughlin-Jastrow part is the ``charge" part of the wave function, while the rest is the ``neutral" part. In $\Psi_{exc}$ the neutral part is the condensate of pseudospins. In $\Psi_{FL}$ the neutral part contains the Slater determinant which introduces the complex conjugate coordinates $\bar z$. Thus $\Psi_{FL}$ in general has a non-zero projection on many Landau levels and not just the LLL. If $\Psi_{FL}$ is to be used a LLL trial wave function, in general it must be explicitly projected via a complicated derivative operator $\mathcal{P}_{LLL}$ to make it a function of $z$'s only~\cite{hlr}.  Nevertheless, in the long distance regime with the sample length $L \gg \ell_B$, we may assume that $z$ and $\bar z$'s commute, and thus in this regime the form of the wave function formally decouples into charge and neutral parts.

\subsection{Intermediate distances: a paired phase}\label{sec:intermed}

Given that the QHB ground state in the limits of both small and large $d$ is in a liquid phase, it seems unlikely that a possible intermediate state is a charge-density wave (that breaks the translation symmetry).

A possible way to describe this evolution is to start from the Bose condensate which is smoothly modified by two Fermi condensates that grow as the distance $d$ increases \cite{srm}.
In such a study on the sphere \cite{msr} it was found that the two Fermi condensates pair in the manner of weakly paired BCS $p_x + i p_y$ wave Cooper pairs. (We use the phrase ``weakly paired" to denote Cooper pairs that are not exponentially localized.) Thus, at the end of the evolution of the superfluid state one may expect the following wave function,~\cite{msr}
\begin{eqnarray}
\Psi_{PSF} (\{z^\uparrow \}, \{z^\downarrow \}) & = & \prod_{i<j}^{N/2} (z_i^\uparrow - z_j^\uparrow )^2 \times \prod_{i<j}^{N/2} (z_i^\downarrow - z_j^\downarrow )^2 \nonumber \\
&& {\rm det} (g ({\bf r}_m^\uparrow - {\bf r}_n^\downarrow)), \label{fpairing1}
\end{eqnarray}
to describe the system. We use here and below a shortened form of the wave functions with spin, where spatial coordinates carry spin. It is understood that by combining 
the spatial wave functions with spinor wave functions, and summing with proper sign factors over all ways of choosing which electrons have which spin, we get proper totally antisymmetric wave functions. In Eq. (\ref{fpairing1}) ``${\rm det}$" denotes the antisymmetrized product of $N/2$ pairs of Cooper pair wave functions,
$g ({\bf r}_m^\uparrow - {\bf r}_n^\downarrow)$, where $m,n = 1, \ldots, N/2$ and $m \neq n$.
The Cooper pair wave function $g (\bf{r}_m^\uparrow - \bf{r}_n^\downarrow)$ describes $p_x + i p_y$ pairing, i.e., at small distances it behaves as~\cite{gunnar}
\begin{equation}
g ({\bf r}_1^\uparrow - {\bf r}_2^\downarrow) = (z_1 - z_2)^l h (|{\bf r}_1 - {\bf r}_2 |) \label{sdispairing}
\end{equation}
with $h(0) \neq 0$, and $l = 1$ which fixes the relative angular momentum of the Cooper pair.

Based on the study of Ref. \onlinecite{msr}, other numerical studies (summarized in Sec. \ref{sec:picture}), and our results (Sec. \ref{sec:numerics}), we will adopt the view that an intermediate state does exist and can be modeled as a BCS $p$-wave paired state of composite fermions. The question of whether in the thermodynamic limit we obtain a critical region (phase) or rather a critical point is more subtle because of the limitations of all the existing numerics to small system sizes. Therefore, we will use the term ``intermediate phase" loosely, keeping in mind the possibility of it becoming a critical point in the limit of a very large (clean) system.

In the remaining part of this Section, we will introduce in a formal way a possible alternative description of the intermediate state based on the picture of  composite bosons, which we expect to be relevant in the long distance limit. First, we would like to point out that the most natural long-distance weak pairing behavior of the pair function in Eq. (\ref{fpairing1}) with $l = 1$ would be
\begin{equation}
g ({\bf r}_1^\uparrow - {\bf r}_2^\downarrow) \sim \frac{1}{(\bar z_1 - \bar z_2)}.
\label{pbehavior}
\end{equation}
This coincides with the existence of a BCS description with a gap function $ \Delta_{\bf k} \sim k_x + i k_y $, i.e., $ \Delta_{\bf k}$ is an eigenfunction of rotations in 2D with $l = 1$ [see Ref. \onlinecite{rg}].

This generic behavior may be used to define a suitable candidate for a projected $p$-wave paired composite fermion wave function
valid at all distances in the LLL, 
\begin{eqnarray}
\nonumber \tilde{\Psi}_{PSF} &=& \mathcal{P}_{LLL} \Big\{ {\rm det} \left( \frac{1}{\bar z_m^\uparrow - \bar z_n^\downarrow} \right) \vphantom{\Big\}} \\ 
&& \vphantom{\Big\{ }  \prod_{i<j}^{N/2} (z_i^\uparrow - z_j^\uparrow )^2 \prod_{i<j}^{N/2} 
(z_i^\downarrow - z_j^\downarrow )^2 \Big\}  \label{Pfpairing},
\end{eqnarray}
where $P_{LLL}$ denotes the necessary projection to the LLL (see Appendix \ref{app:proj} for the details of the projection).
Here we confine ourselves to the long-distance regime, i.e., consider Eq.(\ref{fpairing1}) with (\ref{pbehavior}). 
In this regime, the particles may be considered point-like as the relevant probing wavelengths are 
much larger than $\ell_B$. Thus, to obtain the long-distance properties 
we use Eq. (\ref{fpairing1}) with (\ref{pbehavior}), instead of the proper
LLL function [Eq. (\ref{Pfpairing})], assuming as usual that the universal (topological) properties 
are insensitive to the microscopic differences between the two wave functions. 
In the following, we will explain how in this regime of the $p$-wave composite fermon pairing we can also
consider  a composite boson representation, which will enable us to describe the topological features (degenerate ground
states on the torus and quasiparticle content) of the intermediate state. 

Thus the model wave function for the state into which the superfluid evolves, in the long-distance limit when $z$ and $\bar z$ commute, is
\begin{eqnarray}
\Psi_{PSF} =  \prod_{i<j}^{N/2} (z_i^\uparrow - z_j^\uparrow )^2 \prod_{i<j}^{N/2} (z_i^\downarrow - z_j^\downarrow )^2 {\rm det} \left( \frac{1}{\bar z_m^\uparrow - \bar z_n^\downarrow} \right). \label{fpairing}
\end{eqnarray}
Using the Cauchy identity, we can rewrite the wave function in the following form
\begin{eqnarray}
\Psi_{PSF} =  \prod_{i<j} (z_i - z_j ) {\rm det} \left(\frac{1}{z_m^\uparrow - z_n^\downarrow} \right) {\rm det} \left( \frac{1}{\bar z_m^\uparrow - \bar z_n^\downarrow} \right).
\label{twodet}
\end{eqnarray}
Again, at long distances the charge and neutral part formally decouple, and the neutral part can be viewed as a wave function of some underlying composite boson quasiparticles. Furthermore, the neutral part is the product of two determinants, and a careful conformal field theory (CFT) analysis \cite{mvm_zp,cf} will allow for an existence of a
branch of gapless excitations (see Appendix A) on top of this non-chiral state.

These formal arguments also suggest that there is a possibility to interpret the intermediate phase in the QHB in terms of bosonic quasiparticles that correlate in a special way via Eq.~(\ref{twodet}). From Eq.~(\ref{twodet}) (see also Appendix A) we can read off that the one-particle exciton (or composite boson) correlation function is
\begin{equation}
\lim_{r\to\infty}\langle\psi^{\dagger}_{\uparrow}(\mathbf{r}) \psi_{\downarrow}(0) \rangle \sim \frac{1}{r^2}.
\end{equation}
Therefore, instead of the long-range order that we had in Eq.~(\ref{ferro}), in this ground state we have quasi long-range order. Indeed, in the finite-size study performed in Ref.~\onlinecite{msr}, it was noticed that the excitonic superfluid order parameter, i.e. $\langle\psi^{\dagger}_{\uparrow}(\mathbf{r}) \psi_{\downarrow}(0) \rangle$, persists in the state with no correlated bosonic quasiparticles in the manner of Eq.~(\ref{ferro}) (i.e., the so-called ``111" state~\cite{halp}).

It follows that these critical bosons [Eq.~(\ref{twodet})] may exist in a region between a ferromagnetically ordered and a
completely disordered phase, and thus we may expect a gapless mode and compressible behavior in the neutral sector. This is supported by exact diagonalization studies  on the sphere \cite{gmshs}, but also by the detection of ordered (correlated) behavior in early experiments \cite{spie}, which were performed at distances where we expect the physics of the intermediate state to be relevant.

But what makes the intermediate region so attractive, as we will show in greater detail by exact diagonalization studies on torus [Sec.~\ref{sec:numerics}], is the existence of competing ground states at momenta that diverge in the thermodynamic limit, and thus define different sectors of the system. This signals a topological structure that coexists with gapless behavior. A convenient way of describing various sectors, as we will argue, is to take the composite boson view of the ground state pairing correlations. Then Bose  condensates under different boundary conditions \cite{rg,gth} can be identified with the competing ground states from exact diagonalization studies on the torus.

As usual in topological systems, the torus ground states can be connected with quasiparticles of the topological phase under consideration. We expect this also to be valid in our gapless system, and by Bose condensing in two different pseudospin states, we will be able to connect different ground states with meron spin-textures. This opens a door for a possibility that merons -- fractionally charged vortices of the $XY$ ordered state \cite{moon} -- exist as free (deconfined or not bound in pairs) quasiparticles in the intermediate state.

The first exact diagonalization study on the torus that pointed out to the existence of the competing ground states with momenta that diverge in the thermodynamic limit was done by Park \cite{park} (see Appendix \ref{app:park} for a brief review of the relationship between Ref.\onlinecite{park} and our work). In Ref. \onlinecite{park} a spiraling structure in the competing ground states was identified, and it was argued that the intermediate phase possesses a pseudospin spiral order. Therefore, the competing states
were interpreted as new developing ground states that replace the $XY$ ordered state (valid for smaller distances) due to a (classical) symmetry-breaking ordering. On the other hand, in this work we give numerical and analytical reasons to consider an intermediate state with richer topological content than the $XY$ ordered state (which has a simple integer quantum Hall effect characterization). In the intermediate phase, due to \emph{topological} order, new quasiparticles emerge and define additional ground states on the torus which have a spiraling structure.  In this way our work suggests a way to reconcile and unify the views of the intermediate phase based on the previous numerical approaches, especially Ref. \onlinecite{msr} and Ref. \onlinecite{park}.

\section{The formation of the intermediate state: a physical picture}\label{sec:picture}

One may naively estimate that the transition to a disordered phase occurs around $d\sim \ell_B$, and this estimate is in agrement with several results in the literature that we summarize here. (i) The calculated treshold for the disappearance of the long-range correlated composite bosons and the extrapolated estimate for the vanishing of the long-range order parameter in Ref.~\onlinecite{msr}. (ii) The change in the DMRG spectrum computed in Ref.~\onlinecite{nsdy}, i.e., a ``change in the nature of excited state" with a ``sudden decrease in the excitation gap". (iii) An abrupt change in the velocity of the Goldstone mode in Ref.~\onlinecite{gmshs}. (iv) Exact diagonalization (ED) studies (EDs) presented here as well as in Ref.~\onlinecite{park}, showing the level crossing for the first excited state around $d\sim \ell_B$.

The physical picture of the evolution of the bilayer is the following: in the ordered $XY$ phase we have locally equal probability to find an up and down exciton. As we increase the distance between the layers, the intra-layer Coulomb interaction becomes dominant and we need to place the up excitons, and separately down excitons, further apart. One natural resolution would be an Ising order (charge density wave), but another viable possibility is to slowly disturb the long-range ordering and create spin-textures as ground state(s). In the magnetic description of Ref. \onlinecite{bmd}, the tendency to create spin-textures may be recognized by the presence of frustrating next-nearest-neighbor Ising couplings. Thus the magnetic (bosonic) description of the intermediate phase may be useful along with the fermionic description presented in Ref. \onlinecite{msr}.

\section{Exact diagonalization results on the torus}\label{sec:numerics}

In this Section we present results of exact diagonalization calculations performed on the torus geometry. We model a finite system of $N$ electrons moving on a surface spanned by vectors $\mathbf{L}_1$ and $\mathbf{L}_2$. We fix the filling fraction to be one, which leads to the condition $|\mathbf{L}_1 \times \mathbf{L}_2 | =  2\pi \ell_B^2 N$. The focus of our investigation is on the liquid phases, hence we consider an isotropic torus geometry, i.e., the length of the two vectors is the same (aspect ratio equal to one). Additionally, we can also vary the angle $\theta$ between the two vectors $\mathbf{L}_1$ and $\mathbf{L}_2$. Unless otherwise specified, we fix $\theta=\frac{\pi}{2}$, corresponding to centered rectangular symmetry.

\begin{figure}
\centering
\includegraphics[width=0.95\linewidth]{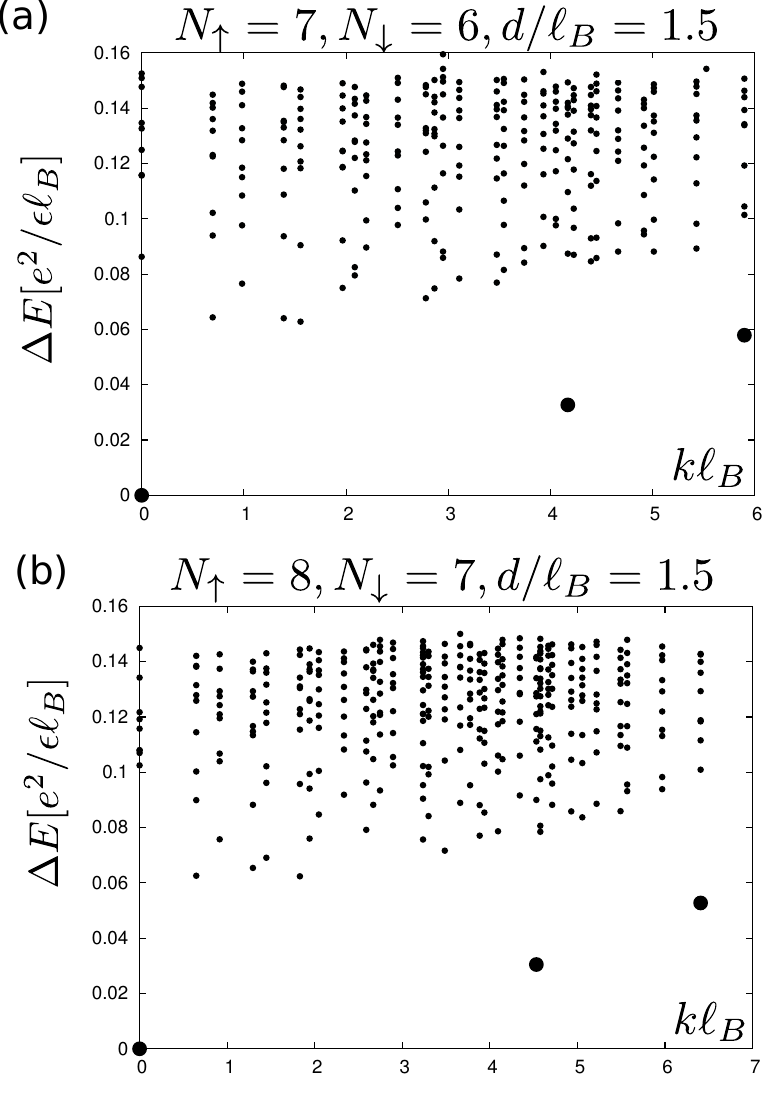}
\caption{\label{Fig:spectrum1315} Energy spectrum versus momentum $k\ell_B$ in the odd sector on the torus. Panel (a) shows the spectrum for $N=13$ electrons, while panel (b) is for $N=15$. In both cases the aspect ratio $r=1$, $\cos\theta=0$ and bilayer distance $d=1.5\ell_B$. Levels are plotted relative to the corresponding ground state.}
\end{figure}

The interaction between electrons consists of an intra-layer Coulomb potential, given by the Fourier transform
\begin{equation}
V_{intra} (q) = \frac{2\pi}{q},
\end{equation}
and the inter-layer Coulomb repulsion which is screened by the bilayer distance
\begin{equation}
V_{inter} (q) = \frac{2\pi}{q} e^{-qd}.
\end{equation}
The interaction potentials are made appropriately periodic (i.e., $q$ assumes values $2\pi/L_{1,2}$ times an integer) to satisfy the periodic boundary conditions.

\begin{figure*}[ttt]
  \begin{minipage}[l]{\linewidth}
\includegraphics[width=\linewidth]{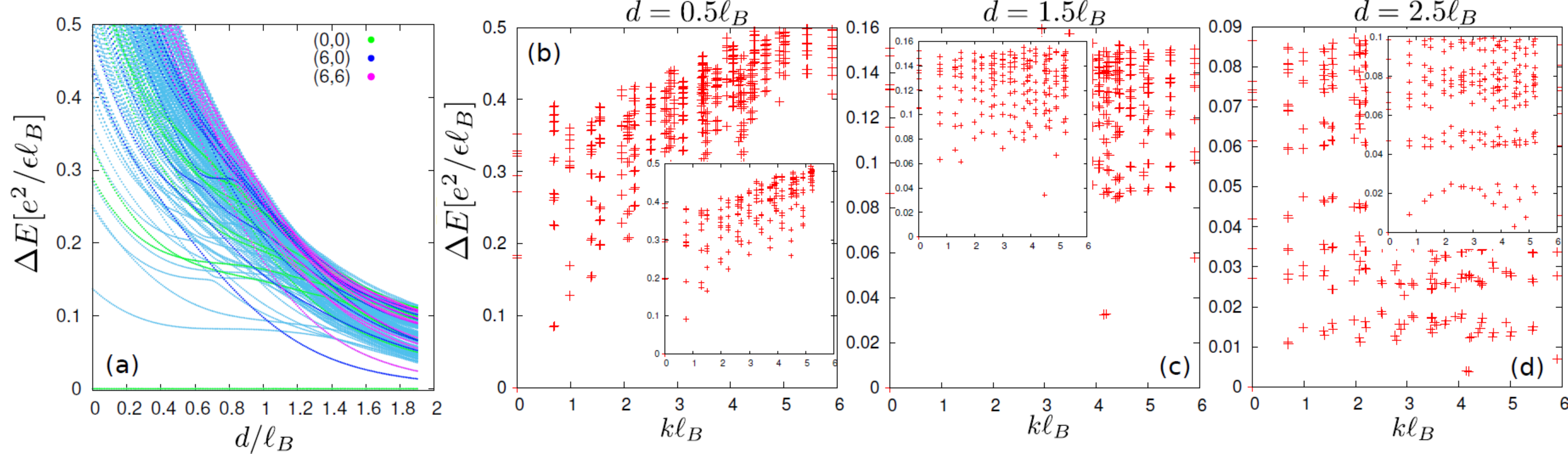}
  \end{minipage}
\caption{\label{Fig:spectrum13dk} (Color online) (a) Evolution of the energy spectrum in the odd sector on the torus as a function of bilayer distance $d/\ell_B$. Levels are plotted relative to the ground state at each $d$ and quoted in units of $e^2/\epsilon\ell_B$. Levels belonging to the three high-symmetry sectors have been designated by colours indicated in the legend. (b)-(d) Energy spectrum versus momentum $k\ell_B$ in the odd sector on the torus for $d=0.5\ell_B, 1.5\ell_B, 2.5\ell_B$. In all cases $N=13$, $\cos\theta=\pi/2$, and aspect ratio is set to $r=0.99$ to slightly break the discrete rotation symmetry. Levels are plotted relative to the corresponding ground state for each value of $d$. Insets: Hexagonal geometry ($\theta=\pi/3$ and $r=1$).}
\end{figure*}

After specifying the geometry and the interaction potential, the basic procedure of exact diagonalization is to represent the Hamiltonian of the system as a large finite matrix and diagonalize it on a computer. In order to extract physical information about the energy spectrum and system's eigenstates, it is essential to label them with good quantum numbers. In this case, the complete symmetry classification was first achieved by Haldane~\cite{haldane_torus}; the primary quantum number that emerges from such an analysis is the two-dimensional momentum that characterizes the invariance under magnetic translations. This momentum is generically conserved for any translationally invariant interaction, and we will use it below to label the eigenstates of the system (see Ref. \onlinecite{bern} for a recent overview of the many-body translation symmetry classification in the magnetic field).

In Figs. \ref{Fig:spectrum1315} and \ref{Fig:spectrum13dk} we present ED results on the torus of the bilayer system with odd numbers of particles. In Fig. \ref{Fig:spectrum1315} we show the energy spectrum versus conserved momentum $k \ell_B$ in the systems with $N = 13$ (a) and $N = 15$ electrons (b). In this figure, the bilayer distance is fixed at $d=1.5\ell_B$, i.e., in the middle of the transition region. Because the total number of particles $N$ is odd, we restrict to the the largest sector of the Hilbert space with $N_\uparrow=(N+1)/2$ and $N_\downarrow=(N-1)/2$.

More generally, the evolution of the entire spectrum as the bilayer distance is varied in presented in Fig. \ref{Fig:spectrum13dk}(a). In Fig. \ref{Fig:spectrum13dk} (b)-(d) we show several snapshots of the energy spectrum in Fig. \ref{Fig:spectrum13dk}(a) plotted as a function of momentum. We choose three values of $d=0.5\ell_B, 1.5\ell_B, 2.5\ell_B$ that tentatively correspond to three different regimes: the 111 state, intermediate region, and the onset of decoupled composite Fermi liquids. The unit cell in Figs. \ref{Fig:spectrum13dk}(b)-(d) is approximately the square, and insets show the same results for the hexagonal geometry.

What we can see in both geometries [Figs. \ref{Fig:spectrum13dk}(b)-(d)] as well as Fig. \ref{Fig:spectrum1315}, is that at distance $d = 1.5 \ell_B$ two distinct states ``fall" from the excited (continuum) part of the spectrum: for example, the state around $k \sim 4\ell_B^{-1}$ and another one at $k \sim 6\ell_B^{-1}$ in Fig. \ref{Fig:spectrum13dk}(c). The state with the momentum $k \sim 4\ell_B^{-1}$ was first pointed out in the study of Park \cite{park}. Our results suggest that this state evolves hand in hand with another one at momentum $k \sim 6\ell_B^{-1}$ in Fig. \ref{Fig:spectrum13dk}(c). The presence of these two states is robust and does not depend on system size, for example a larger $N=15$ system shown in Fig.\ref{Fig:spectrum1315} contains the same doublet of states. We notice that these ``special" states compete with the ground state at $\mathbf{k} = 0$ as $d$ becomes larger.

The two special states in fact each comprise a family of 4 states connected by symmetry operations (simple translations and inversion). Thus, in Fig.~\ref{Fig:spectrum1315} we find a set of 4 states that correspond to the lower energy excitation (with $k\sim 4\ell_B^{-1}$):
\begin{eqnarray}\label{state1}
(N_{\uparrow},0), \; (N_{\downarrow},0), \; (0,N_{\uparrow}) \; {\rm and} \; (0,N_{\downarrow})
\end{eqnarray}
in units of $2\pi/L$ where $L=\sqrt{2\pi \ell_B^2 N}$. This set of states is followed by the excitation at
\begin{eqnarray}\label{state2}
(N_{\uparrow},N_{\uparrow}), \; (N_{\downarrow},N_{\uparrow}), \; (N_{\uparrow},N_{\downarrow}) \; {\rm and} \; (N_{\downarrow},N_{\downarrow})
\end{eqnarray}
with a slightly higher energy (with momentum $k\sim 6\ell_B^{-1}$). We note that we have identified such states in all odd systems we studied, e.g. $ N = 9, 11, 13 ,$ and $15$.

\begin{figure}[htb]
\centering
\includegraphics[width=0.95\linewidth]{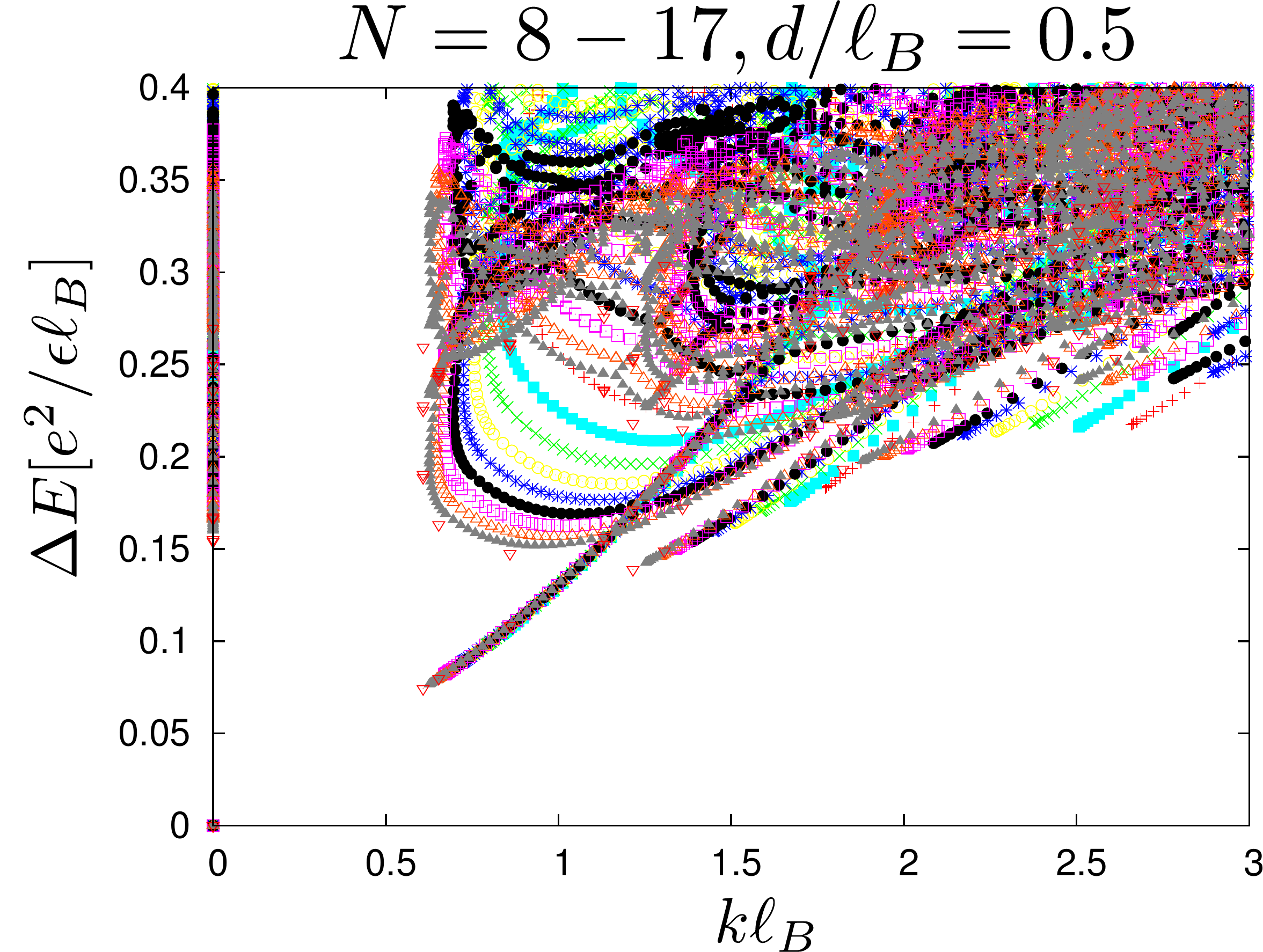}
\caption{\label{Fig:goldstone} (Color online) Goldstone mode for the bilayer distance $d=0.5\ell_B$. Quasicontinuous $E(k)$ energy spectrum is obtained by varying the torus angle $\theta$ for a fixed aspect ratio $r=1$. Data is shown for all system sizes between $N=8$ and $N=17$ electrons in total. Levels are plotted relative to the corresponding ground state for each $\theta$ and $N$.}
\end{figure}

\begin{figure}[htb]
\centering
\includegraphics[width=0.95\linewidth]{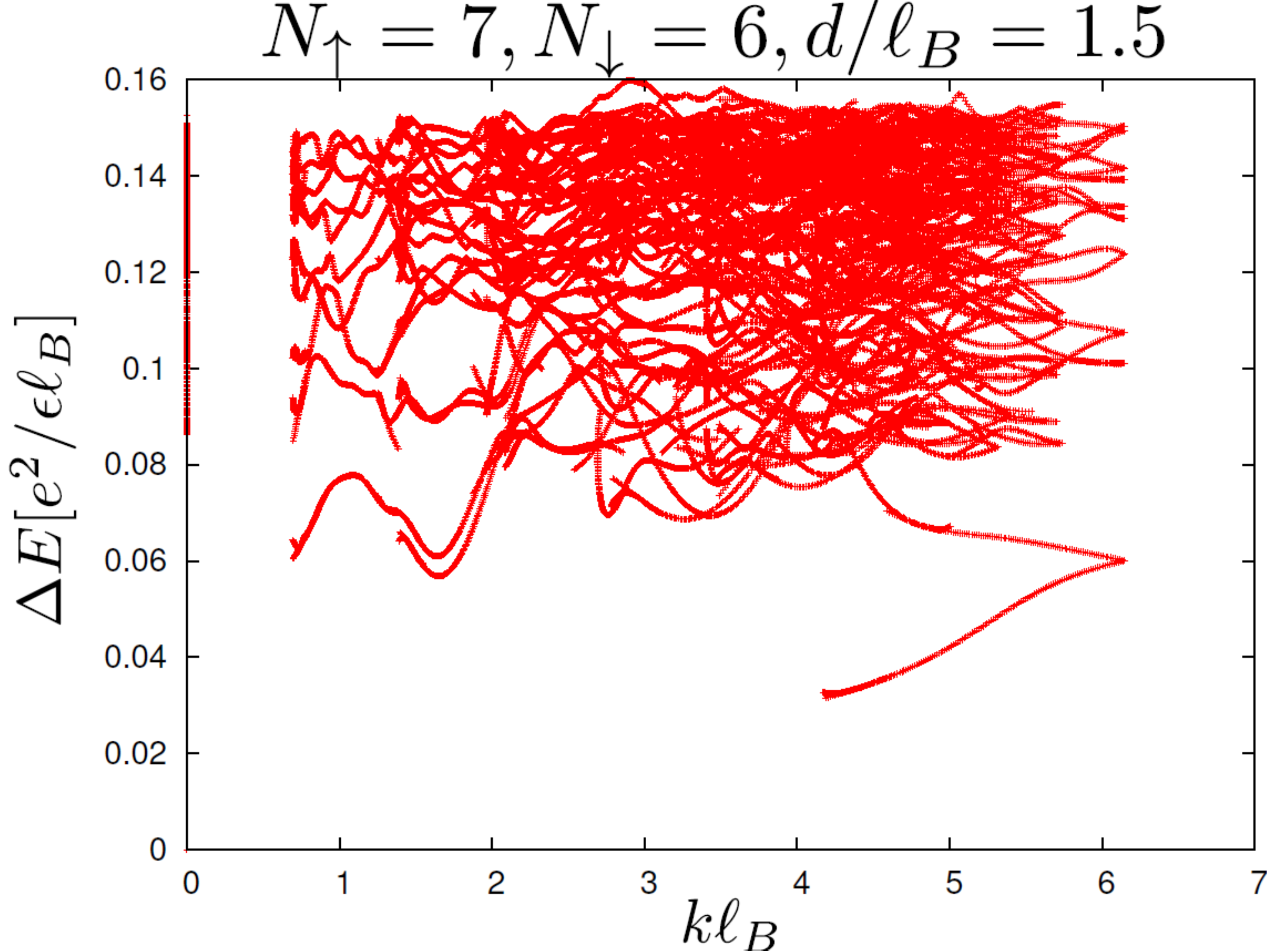}
\caption{\label{Fig:spectrum13angle} (Color online) Quasicontinuous $E(k)$ energy spectrum obtained by varying the torus angle $\theta$ for a fixed aspect ratio $r=1$. Total number of electrons is $N=13$ and bilayer distance is $d=1.5\ell_B$. Levels are plotted relative to the corresponding ground state for each $\theta$. Levels $(N_\uparrow,0)$ and $(N_\uparrow, N_\uparrow)$ belong to the same excitation branch.}
\end{figure}

Thus we may conclude that at distances where we may expect the intermediate phase, by examining the  torus results we find competing ground states at momenta that diverge in the thermodynamic limit. This may be indicative of the presence of Bose condensation and spin textures, but requires additional analysis.

Studies of the gapless Goldstone mode, e.g. the numerical one on the sphere~\cite{gmshs} and the experimental detection in Ref. \onlinecite{spie}, suggest that even though the numerical work, as summarized in Sec.~\ref{sec:numerics}, points to the formation of the disordered state around $ d \sim \ell_B$, a linear gapless mode persists to exist in this phase. Thus although we may notice ``a small gap structure" in Figs. \ref{Fig:spectrum13dk}(d) after $ d \sim \ell_B$, it is likely that it does not represent a paramagnetic phase. Because of the finite size effects and geometry, this ``weak Goldstone" mode of, as we will argue, critical bosons, is distorted. Despite the distortions, in the hexagonal geometry at $d =  2.5 \ell_B$, we can see clearly, as we explain in the following section, consequences of the physics of critical bosons in the intermediate state: we find three (almost) degenerate ground states (families with members connected with symmetry operations) and discernible gapless modes that have beginnings in these states.

In order to examine in more detail the question of the existence and description of gapless modes by EDs, energy spectrum was obtained by varying  the torus angle $\theta$. In this way, as we sweep $\theta$ continuously between $\pi/2$ and $\pi/3$, we obtain many different resolutions of momenta $\mathbf{k}$, which produces a quasicontinuous energy spectrum. In Fig. \ref{Fig:goldstone} we perform such a calculation in the ordered phase ($d=0.5\ell_B$), where we find clear evidence for a linear gapless mode. In Fig. \ref{Fig:spectrum13angle} we perform a similar calculation for $d=1.5\ell_B$, i.e. in the intermediate region. Here it is strongly suggestive that the two families of states in Eqs. (\ref{state1}) and (\ref{state2}) indeed evolve together as the system is adiabatically deformed, and likely belong to the same gapless excitation branch. The distortions of a finite system do not give us a rigorous proof of the gapless mode, but we can expect that if such a mode exists it will appear in the manner produced by the probe. Thus with the distortion we can just produce a state at the ``distorted momentum", or we can access the energy of the state belonging to the excitation branch in the 2D momentum space, at missing (undistorted) momentum in the finite size system.

\begin{figure}[htb]
\centering
\includegraphics[width=0.95\linewidth]{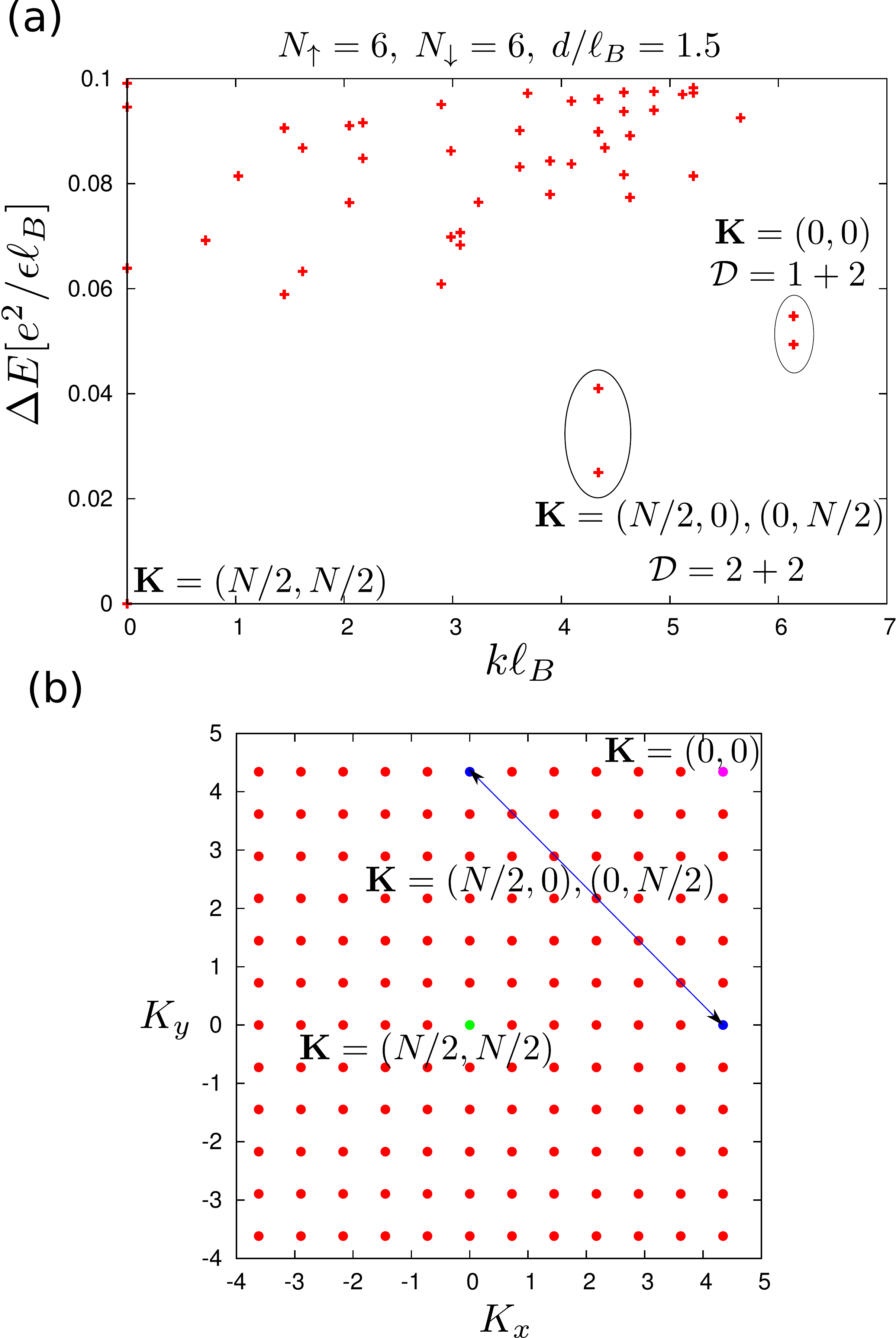}
\caption{\label{Fig:spectrum12} (Color online) (a) Energy spectrum versus momentum $k\ell_B$ in the even sector on the torus for $N=12$ electrons in total. Aspect ratio $r=1$, $\cos\theta=0$ and bilayer distance $d=1.5\ell_B$. Levels are plotted relative to the corresponding ground state at $(N/2,N/2)$. Ellipses denote groups of levels in momentum sectors $(0,0)$ and $(N/2,0)$. Total degeneracy inside each ellipse is denoted by $\mathcal{D}$. (b) The Brillouin zone for $N=12$ corresponding to the square unit cell.}
\end{figure}

In the even sector ($N$  even) of the EDs on torus, Fig. \ref{Fig:spectrum12}, we can similarly recognize the competing ground states with momenta that diverge in the thermodynamic limit. We will postpone the  discussion of these results to Sec.~\ref{sec:competing} and Sec.~\ref{sec:topo}. Before that, in the same section, it will be explained how ground states at $k \neq 0$ emerge from the physics of critical bosons.

\section{A state of critical bosons at intermediate distances}\label{sec:boson}

\subsection{The intermediate state and critical bosons}

Though from the torus results (Fig. \ref{Fig:spectrum1315} and Fig. \ref{Fig:spectrum12}) we cannot claim the
ideal degeneracy of the competing states, their configuration is robust and characterizes the intermediate region. (We have verified this in systems of size up to 15 electrons in the odd sector, and up to 14 electrons in the even sector.)
Therefore, the systems analyzed are likely near a critical point, line, or perhaps even a region \cite{msr} in the parameter space where there is ideal degeneracy, implied by the topological structure of the model function [Eq.~(\ref{fpairing}) or Eq.~(\ref{twodet})].

In the following we explain how the ``extra" competing ground states at non-zero momenta can be interpreted as pairs of merons (charged vortices of the superfluid state at small $d$), whose spiraling period is commensurate with the length(s) of the system. These spiraling configurations were considered in the work of Park \cite{park}, but here we would like to point out that on the torus they can be viewed as a consequence of the physics of critical bosons. In this way we will relate the physics of long spiraling on the torus and critical (composite) bosons to the existing description in terms of $p$-paired states of composite fermions (see Introduction and Ref. \onlinecite{msr}).

To capture the critical correlations of bosons, in the following we will consider bosons with the critical BCS pairing function $g(r)\sim1/r^2$, i.e., $g(k)\sim - \ln |k|$ in momentum space. Thus we assume that the dominant contribution in Eq.~(\ref{twodet}) are pairwise correlations and that the state can be considered as a BCS correlated state of bosons (where the correlations are expressed in terms of a \emph{permanent} and not the more usual determinant).

The assumption of only pairwise correlations is by no means self-evident, and there are two
reasons why it is justified: (a) in our system there is no real charge - pseudospin separation or, at least we have to allow for such a possibility, and thus ``halves" of the (fused) quasiparticles [in Appendix \ref{sec:appA}, the quasiparticle in Eq.(\ref{qp2}) is a product of two fused ``halves"] in the pseudospin sector are also allowed and should be captured in the critical boson description, and (b) the second reason is heuristic: by allowing the reduction to pairwise correlations we can describe the ground states on the torus. The pairing allows the constructions of ``halves", e.g. ``halving" the Laughlin quasihole into two non-Abelions in the Moore-Read state \cite{mr}, and by keeping only pairing correlations
the degrees of freedom that would be forgotten (suppressed) otherwise will be released.

Thus we assume that the  low-energy states of the intermediate phase are in one to one correspondence to the states of $g(r) \sim 1/r^2$ BCS paired critical bosons.  Now we proceed similarly to the case of permanent pairing $g(r) \sim 1/z$ in Ref. \onlinecite{gth}. In the effective BCS description the occupation number of bosons at momentum ${\bf k}$ is
\begin{equation}
\langle n_{{\bf k}} \rangle = 2 \frac{|g({\bf k})|^2}{1 - |g({\bf k})|^2},
\label{occupation}
\end{equation}
where we note the minus sign with respect to the fermionic case. To maintain the positivity of the expression with $g({\bf k})= - \lambda \ln{k}$, where $\lambda$ is a constant with respect to ${\bf k}$,  we must have  ${\bf k} \neq {\bf 0}$. This means we must be in the antiperiodic sector on the torus along at least one direction ($x$ or $y$). Further, we also must have $ \lambda \sim 1/\ln{k_{min}}$ because $k_{min} \sim \pi/L \sim 1/N$. Finally, because $\lambda \rightarrow 0$ as $N \rightarrow \infty$, i.e., no pairing  in the thermodynamic limit, we can only have Bose condensation (at $k_{min}$).

Let us assume (as in Ref. \onlinecite{gth}) that we are in the $(-,+)$ sector, i.e., antiperiodic in $x$ direction and periodic in $y$ direction. We assume an even number of particles (i.e., that we are in the even sector), and occupy $(+ \pi/L,0)$ and $(-\pi/L,0)$ with equal probability, i.e.,
\begin{equation}
b_{k \uparrow} = \sqrt{N/2}\; \delta_{k, k_{min}} \;\;\;\; {\rm and}\;\;\;\; b_{k \downarrow} = \sqrt{N/2}\; \delta_{k,- k_{min}},
\end{equation}
where $ k_{min} = \pi/L$, and $b_{k \sigma}, \sigma = \uparrow, \downarrow$ denote second-quantized bosonic operators.

Thus in the neutral sector each particle would have a description in the form of a two-component spinor:
\begin{equation}
\left[\begin{array}{c}  e^{i Q x} \\
                        e^{-i Q x}             \\
\end{array} \right]
\label{description}
\end{equation}
where $Q = \pi/L$. On the other hand, in Park's states, Eq. (9) in Ref. \onlinecite{park}, we have the effective description:
\begin{equation}
\left[\begin{array}{c}  e^{i 2 Q x} \\
                        1            \\
\end{array} \right],
\label{Pdescription}
\end{equation}
that successfully reproduces the momenta that we see in EDs on the torus.
We may notice that the difference comes from the necessity to multiply the neutral part in Eq.(\ref{description}) by $\exp(i Q x)$ in order to ``glue" the charge (periodic) and neutral (antiperiodic) sector, and stay in the periodic sector.  This ``gluing" effectively brings non-zero momentum to the state: $K_x = N Q = N \pi/L$. In Park's picture the momentum count results from excitons turning into dipoles (not locally neutral particles), and each exciton acquiring the momentum $\tilde{k} = 2 \pi/L$. Thus, at small length scales, charge and spin are not separated, and this corresponds to the requirement of gluing, i.e., staying in the charge-periodic sector by an extra connection between charge and spin sectors. At the end, we have a spiral  winding with $ q = 2 Q = 2 \pi/L$, i.e., the spins wind exactly once over the length of the system, $L$, in $\hat{x}$ direction. Thus, due to the intertwined spin and charge degrees of freedom at small length scales, we obtain ground states at large non-zero momenta.

But there is another important relation of the ``gluing" procedure with what we can infer from EDs on torus. The description of the intermediate phase that is based only on the neutral sector is incomplete. As usual in FQH systems, the degeneracy of the ground states on the torus can be connected with ``large gauge transformations", which are flux insertions of gauge (internal) fields through the two holes of the torus \cite{rg}. This is equivalent to changing the boundary conditions, but also to the creation of particle - antiparticle pairs. These transformations can create large momenta that characterize the degenerate ground states, or alternatively different sectors of the theory [see, for example, a recent discussion in Ref. \onlinecite{4r}]. Thus we also have to take into account the charge part of the theory in order to describe the ground state.

Since the work of Moon \emph{et al.} \cite{moon}, we know that vortex excitations in the exciton superfluid phase are charged -- they are merons of unit vorticity $(\pm 1)$ and charge $\pm (1/2) e$. We may then question the CFT approach for constructing excitations restricted to the neutral sector [see Appendix \ref{sec:appA}], or in other words whether the excitations obtained by Eq. (\ref{qp2}) comprise all possible vortex excitations in the intermediate critical phase.

In answering this question we may start from the model function for Park's states [Eq. (8) in Ref. \onlinecite{park} or Eq.~\ref{cmeron} here]. This construction represents two merons with opposite charge and located at two opposite boundaries [compare with Eq.(135) in Ref. \onlinecite{moon}]. On the disk, this wave function represents one meron at the center of the disk and the other at the infinity. (This is somewhat similar to the construction of the non-Abelian quasiparticles in the case of Moore-Read state which also come in pairs \cite{mmnr}.) In this way, because the meron boundary constructions define the ground states, we may expect that merons form the quasiparticle content of the critical theory. Charge and spin are not separated in these constructions that describe the ground states of the critical phase. Thus it was necessary to include the ``gluing" when discussing the ground state from the point of view that treated the charge and spin sector separately.

More simply, on the disk with a fixed number of up and down particles we model a meron as
\begin{equation}
\prod_i (z_i^{\sigma} - w) \Psi_{gs}(z^{\uparrow},z^{\downarrow} ), \;\; \sigma = \uparrow, \downarrow,
\label{meron}
\end{equation}
where $\Psi_{gs}$ and $w$ denote the ground state and the position of the meron respectively. Now we see the need for the square root factors, $\prod_i \sqrt{(z_i - w)} = \prod_i \sqrt{(z_i^{\uparrow} - w)} \prod_i \sqrt{(z_i^{\downarrow} - w)}$, in the charge part (``a charge boost") to relate to the expression in Eq.(\ref{qp2}), which represents a neutral combination of two merons.

\subsection{Competing ground states on the torus and critical bosons}\label{sec:competing}

To understand better the competing (ground) states present in EDs of the QHB on the torus in the intermediate region, we will consider in more detail the odd and even sectors. As in the previous subsection, we will relate the description of states in terms of Bose condensates at non-zero momenta with meron-antimeron constructions. For further discussion concerning the validity of the description by these states, see Appendix~\ref{Slowlying}.

In the odd sector, for the periodic boundary conditions in both directions, i.e., in the $(+, +)$ sector, we have simply Bose condensation in $K = (0,0)$ momentum. The relevant construction in the odd  $(-,+)$ sector, i.e. antiperiodic in the $x$ direction and periodic in the $y$ direction, is
\begin{equation}
\prod_{i=1}^{N-1}
\left[
  \begin{array}{c}
    \mathrm{e}^{\mathrm{i}\frac{\pi}{L}x_i} \\
    \mathrm{e}^{-\mathrm{i}\frac{\pi}{L}x_i}
  \end{array}
\right]_i
\left[
  \begin{array}{c}
    1 \\
    0
  \end{array}
\right]_N  ,
\end{equation}
where we fixed $N_\uparrow - N_\downarrow = 1$ in the ground state.
Again, we need the gluing procedure with the boost $K=(\frac{N-1}2,0).$
The translation by $-N$ in the $x$-direction followed by inversion lead to the new relevant state $K=(\frac{N+1}2,0).$ Taking also the possibility of the construction along the $y-$axis, we get a 4-fold degeneracy. In the even sector a similar procedure will lead to only the 2-fold degeneracy for the constructions along any one of two directions.

Similarly we can consider the following constructions in the odd $(-,-)$ sector, i.e. antiperiodic in both directions,
\begin{eqnarray}
\prod_{i=1}^{N-1}
\left[
  \begin{array}{c}
    \mathrm{e}^{\mathrm{i}\frac{\pi}{L}(x_i+y_i)} \\
    \mathrm{e}^{-\mathrm{i}\frac{\pi}{L}(x_i+y_i)}
  \end{array}
\right]_i
\left[
  \begin{array}{c}
    1 \\
    0
  \end{array}
\right]_N, \\
\prod_{i=1}^{N-1}
\left[
  \begin{array}{c}
    \mathrm{e}^{\mathrm{i}\frac{\pi}{L}(x_i-y_i)} \\
    \mathrm{e}^{-\mathrm{i}\frac{\pi}{L}(x_i-y_i)}
  \end{array}
\right]_i
\left[
  \begin{array}{c}
    1 \\
    0
  \end{array}
\right]_N,
\end{eqnarray}
and boosts are $K=(\frac{N-1}2,\frac{N-1}2)$ and $K=(\frac{N-1}2,\frac{-N+1}2)$, respectively. By applying the translations by $\pm N$ and inversion, we may conclude that in the odd sector we have a 4-fold degeneracy with $K=(\frac{N\pm1}2,\frac{N\pm1}2).$
On the other hand, a similar construction in the even sector leads to a unique state with $K=(\frac{N}2,\frac{N}2)$.

Thus we can conclude that the prediction for the number of competing ground states in the odd sector, based on the critical boson condensates, is in an agreement with EDs in Section \ref{sec:numerics}, in particular Figs. \ref{Fig:spectrum1315} and \ref{Fig:spectrum13dk} and Eqs. (\ref{state1}) and (\ref{state2}), for the description of the $k \neq 0$ states.

On the other hand, Fig.~\ref{Fig:spectrum12} shows the low lying excitation spectrum in the even sector with $N = 12$ electrons at distance $d = 1.5 l_B$ between the layers. We may notice that with respect to our expectations in the previous considerations (based on meron-antimeron constructions of ground states) there are extra low-lying states. This will be discussed in more detail in the next subsection.

\subsection{Discussion: Topological characterization}\label{sec:topo}

In the previous subsection we could notice that although the composite boson view is convenient for the description of the low-lying states, it is not complete and sufficient for their description. In connection with this, we may comment that the meron construction in Eq. (\ref{meron}) seems more natural if $\Psi_{gs}$ is given by Eq. (\ref{fpairing}) (fermionic representation) than by Eq. (\ref{twodet}) (bosonic representation). In the fermionic case the meron, with the deficit of $\sigma$ (layer) polarization, may be interpreted as a Laughlin quasihole excitation of the condensate of $\sigma$ particles.

In the absence of pairing, represented by the determinant in the composite fermion wave function Eq.(\ref{fpairing}), we could  consider a bosonic system described by Halperin's 220 state \cite{halp} with two components, and merons as  half-flux quantum, spinful quasiparticles. For that system we know the number of the degenerate ground states on the torus; we can consider all possible configurations (including meron ones) between the two edges of a cylinder \cite{mmnr}, and come to the conclusion that there are 4 sectors (4 degenerate states on a torus).  They can be classified as members of the chiral algebra, in which we have the basic sector that we can associate with the identity operation, two sectors due to two kinds of meron - (anti-meron) excitations, and a sector that is associated with the one-flux-quantum excitation i.e. Laughlin quasihole, which can be viewed as a fusion of the two ($\uparrow$ and $\downarrow$) different merons. In this way the four sectors can be viewed as a direct product of two subsystems, each at filling $1/2$ and having two sectors, due to the identity and Laughlin excitation in each system. Symbolically, if we denote the identity by $1$, and $\uparrow$ meron by $m$, and  $\downarrow$ meron by $\bar{m}$, we have a direct product $ (1,m) \times (1,\bar{m}) =\{ (1,1),(m,1), (1,\bar{m}), (m,\bar{m})\}$, i.e., 4 sectors. Having the extra determinant as in Eq.(\ref{fpairing}), beside the (220) factor, will not change the type of the charged excitations that exist in the intermediate phase, i.e., we expect merons. Thus we obtain a total 4-fold degeneracy of the ground states on the torus.

But we have to be aware that our system is far from being a simple bosonic (220) state. There we could consider excitations of a single boson, $\uparrow$ or $\downarrow$, but this does not create an extra sector in the (220) system because the two subsystems, $\uparrow$ and $\downarrow$, are uncorrelated, and everything boils down to the direct product  of $\uparrow$ and $\downarrow$ (sub)system. In the case of the state in Eq.(\ref{fpairing}), because of the interlayer correlations, we may assume that a single fermion excitation, which we will denote by $f$ for $\uparrow$ subsystem and $\bar{f}$ for $\downarrow$ subsystem, is a non-trivial excitation, and may define a sector. Thus we may consider $ (1, m, f) \times (1, \bar{m}, \bar{f})$ as a way to generate all sectors \cite{comment}. But we notice that $(f,\bar{f})$ is equivalent to two $(m,\bar{m})$ excitations, and should not represent an extra sector. Thus, in this way we expect an 8-fold degeneracy in the even sector, and this is corroborated by the results in
Fig.~\ref{Fig:spectrum12}.

This is by no means a usual characterization of a topological phase. The topological characterization is made difficult by the absence of a simple model Hamiltonian for the model wave function [Eq.~(\ref{fpairing}) or Eq.~(\ref{twodet})]. Due to (topologically) highly intertwined (pseudo)spin and charge, we might expect a further reduction of the number of relevant states. Namely, an operation of equivalence may be introduced for two states that are connected by a gapless branch, like in Fig. \ref{Fig:spectrum13angle}.
Nevertheless, we find that the low-lying spectrum, in both even and odd sector is robust and independent of system size, up to $ N = 14 $ particles in the even sector, and up to $N = 15$ in the odd sector, and it is not in the accordance with the naive expectation of 4 degenerate ground states in the even sector that can be found in the previous literature, or as follows from the charge-pseudospin separated picture of the intermediate state.

By comparing the tentative degeneracies on the torus (8 states in the even sector and 9 states in the odd sector) with the predictions in the even and odd sector in section \ref{sec:competing} on the basis of $g(r) \sim 1/r^2$ BCS paired critical bosons, we can conclude that this view is partially successful. We may expect Abelian merons {\em and} composite fermions in the quasiparticle spectrum.

\subsection{Discussion: Meron deconfinement}\label{sec:deconf}


In this subsection we summarize the reasons why meron deconfinement is expected in the intermediate state in the QHB. By ``meron deconfinement" we mean the phenomenon that a meron can exist as a single, gapped excitation of the half-flux quantum and charge $(1/2)e$ in the intermediate phase (as opposed to being in a bound pair).

\begin{figure}[htb]
\centering
\includegraphics[width=\linewidth]{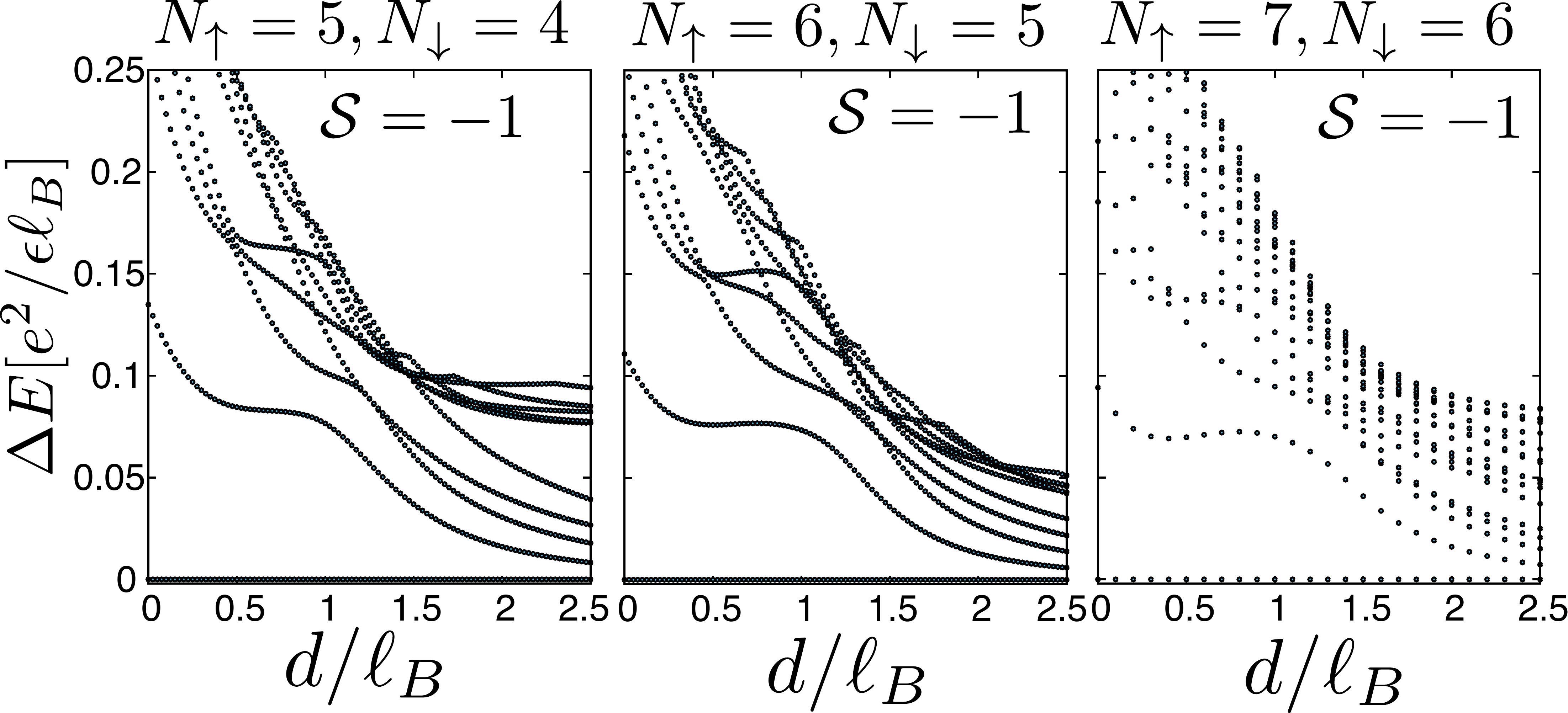}
\caption{\label{sphere} The evolution of the energy spectrum in the odd sector on the sphere as a function of bilayer distance. Data is for $N=9, 11, 13$ electrons in total (left to right), for fixed shift $\mathcal{S}=-1$ corresponding to the 111 state.}
\end{figure}

When we compare the evolution of the QHB with respect to distance, on the sphere [Fig. \ref{sphere}] and on torus [Figs. \ref{Fig:spectrum1315}, \ref{Fig:spectrum13dk}], we see that on the sphere the lowest Goldstone-branch state evolves without level crossing with distance. The calculations on the sphere in Fig.~\ref{sphere} are performed at the fixed shift $\mathcal{S}=-1$, which is the same as that of the 111 state. Note, however, that the decoupled Fermi liquids have a shift of $\mathcal{S}=-2$, therefore the sphere geometry cannot directly capture the transition without introducing some bias towards one of the phases. In Fig.~\ref{sphere} the ground state smoothly changes around $d\sim \ell_B$ due to the change in the correlations, and also there is a drastic change in the calculated slope (velocity) of Goldstone mode according to Ref. \onlinecite{gmshs}, but the first excited state is smoothly connected to the Goldstone branch for any $d$.

In contrast, on the torus [see, e.g., Fig. \ref{Fig:spectrum13dk}(a)] there are level crossings in the low-lying excited spectrum around $d\sim \ell_B$. New states appear with lower energies than the first excited Goldstone-branch state. These states constitute candidate ground states of the intermediate phase. They are defined by merons, as explained previously in Sec. \ref{sec:competing}. Thus merons belong to the quasiparticle content of the new state.

The intermediate state has the interpretation in terms of BCS pairing of composite fermions (Ref. \onlinecite{msr}), and thus similarly to the Moore-Read case we expect half-flux quantum (one flux quantum in the language of superconductivity) excitations.

\section{Conclusions and discussion}\label{sec:conc}

We presented a description of the intermediate region of the QHB in terms of critical bosons based on detailed exact diagonalizations on the torus. Our study complements earlier results on the sphere \cite{msr} and torus \cite{park} geometry, where an intemediate phase was identified at bilayer distances $d\approx 1.5\ell_B$. These studies have pointed out that the phase involves pairing of composite fermions, and has pseudospin spiral order. Our work shows that this state can also be related to critical bosons and is likely gapless. The critical condensates at finite wave vectors of the intermediate phase on the torus can be related to the existence of merons as the lowest lying charged quasiparticles, i.e., to meron deconfinement. The topological ordering in the intermediate region with meron fractionalization allows a smooth transition (ordering) in the adjusted phases in which merons are confined in pairs; as vortex excitations in the $XY$ ferromagnet for small d, and, for large d, as bound pairs
they make neutral fermions.

We caution that the critical boson representation employed here may be less powerful than similar descriptions of the more conventional FQH states where charge and pseudospin (neutral degrees) are not so intertwined. In our case, the natural representations in terms of (composite) bosons (exciton superfluid) and (composite) fermions (composite Fermi liquids) hold in the limiting cases of small and large distances. However, we showed that most of the low-energy features of the intermediate state can be conveniently explained in the picture of critical bosons.

Assuming the intermediate state develops into a critical point in the thermodynamic limit, the QHB system may offer a realistic application of the ideas of deconfined criticality \cite{sv}. Moreover, the presented QHB description may also explain the phenomenon of ``meron loosening" in experiments \cite{dz}. Before invoking any disorder effects, we may assume that merons are liberated (not bound in pairs with opposite layer indexes), and equally contribute to inter and intra dissipative processes that lead to the breakdown of the quantum Hall effect. As argued in Ref. \onlinecite{ws}, the quasiparticles with one-half of the elementary charge are natural (confined in dipoles) constituents of the underlying composite fermions in the Fermi-liquid-like state (at filling one-half), due to the particle-hole symmetry. In our case these are merons (of the same layer index), which are brought to the existence in the intermediate state, before the two separate Fermi-liquid-like states (each at filling one-half) set in.

\appendix
\section{The bosonic interpretation}\label{sec:appA}

We can rewrite the neutral part in Eq.~(\ref{twodet}) as
$$
\prod_{i<j} (z_i - z_j ) \times \frac{\prod_{i<j} |z_i^\uparrow - z_j^\uparrow |^2 \times \prod_{p<q} |z_p^\downarrow - z_q^\downarrow |^2}{\prod_{k,l} |z_k^\uparrow - z_l^\downarrow |^2}.
$$
Thus, the neutral part in Eq.~(\ref{twodet}) in the CFT language is the correlator of equal numbers of two vertex operators:
$e^{i \phi(z) + i \phi(\bar z)}$ and $e^{- i \phi(z) - i \phi(\bar z)}$ where $\phi(z)$ and $\phi(\bar z)$ denote the holomorphic and anti-holomorphic part of the (non-chiral) bosonic field, $ \phi(z, \bar z) = \phi(z) +  \phi(\bar z)$. The basic correlator is:
\begin{equation}
\langle e^{i \beta  \phi(z_{1},\bar z_{1})} e^{- i \beta
\phi(z_{2},\bar z_{2})} \rangle = \frac{1}{|z_{1} - z_{2}|^{2 \beta^{2}}}
\end{equation}
We expect that the excitations are also represented by vertex operators. For their construction we have two possibilities: $e^{i \delta_1 (\phi(z) + \phi(\bar z))}$ and $e^{i \delta_2 (\phi(z) - \phi(\bar z))}$ and the corresponding conjugates, which in charge-neutral combinations in the correlators with basic particles
($e^{i \phi(z) + i \phi(\bar z)}$ and $e^{- i \phi(z) - i \phi(\bar z)}$) i.e. up and down (neutral) excitons, contribute factors,
\begin{equation}
e^{i \delta_{1}  \phi(w,\bar w)} \rightarrow \frac{\prod_{i}|z_{i}^\uparrow - w|^{2 \delta_{1} }}{\prod_{i}|z_{i}^\downarrow -
w|^{2 \delta_{1} }},\label{qp1}
\end{equation}
and, with the field, $ \theta(z, \bar z) = \phi(z) - i \phi(\bar z)$, we have
\begin{equation}
e^{i \delta_{2}  \theta(w,\bar w)} \rightarrow
\frac{\prod_{i}(z_{i}^\uparrow - w)^{ \delta_{2}}}
{\prod_{i}(z_{i}^\downarrow - w)^{ \delta_{2}}} \times
\frac{\prod_{i}(\bar z_{i}^\downarrow - \bar w)^{ \delta_{2}}}
{\prod_{i}(\bar z_{i}^\uparrow - \bar w)^{ \delta_{2}}}.\label{qp2}
\end{equation}
To ensure the single-valuedness of the ensuing electronic wave function we must have $\delta_{2} = 1/2$. The same argument would allow any rational $\delta_{1}$ including $\delta_{1} = 0$. Therefore the excitation spectrum contains a branch continuously connected with the ground state and we expect that the state in (\ref{twodet}) supports a branch of gapless excitations. (If we allow only $\delta_{1} = 1/2$ we would have the quasiparticle content of the deconfined phase of $\mathbb{Z}_2$ gauge theory.) Thus the state (\ref{twodet}) is a critical state with the quasi-long-range order,
\begin{eqnarray}
&&\lim_{|z_1 -z_2| \rightarrow \infty} \langle \Psi_\uparrow^{ex}(z_1, \bar z_1) \Psi_\downarrow^{ex}(z_2, \bar z_2)\rangle = \nonumber \\
&& \langle e^{i  \phi(z_{1},\bar z_{1})} e^{- i
\phi(z_{2},\bar z_{2})} \rangle = \frac{1}{|z_{1} - z_{2}|^{2}} \nonumber \\ \label{quasi}
\end{eqnarray}
among up and down excitons.

\section{Projection to the LLL}\label{app:proj}

In this Appendix we explain the projection to the LLL \cite{jach} that defines
the LLL projected wave function of $p$-wave paired composite fermions in
Eq.(\ref{Pfpairing}). First we expand and rewrite Eq.(\ref{Pfpairing}) as
\begin{eqnarray}
\nonumber \tilde{\Psi}_{PSF} &=& \mathcal{P}_{LLL} \Big\{ \sum_{\sigma \in S_{N/2}}  
\frac{1}{(\bar z_1^\uparrow - \bar z_{\sigma(1)}^\downarrow)\cdots 
 (\bar z_{N/2}^\uparrow - \bar z_{\sigma(N/2)}^\downarrow)}  \vphantom{\Big\}} \\ 
 && \vphantom{\Big\{} \Psi_{220}(z_1^\uparrow,\ldots, z_{N/2}^\uparrow, z_1^\downarrow, \ldots, z_{N/2}^\downarrow) \Big\}  \label{Pexpand},
\end{eqnarray}
where by $\Psi_{220}$ we denoted the part with the Laughlin-Jastrow correlations,
\begin{eqnarray}
\Psi_{220} =
 \prod_{i<j}^{N/2} (z_i^\uparrow - z_j^\uparrow )^2 \prod_{i<j}^{N/2} 
(z_i^\downarrow - z_j^\downarrow )^2 . 
\end{eqnarray}
To understand the projection, we consider an arbitrary permutation $\sigma$
under the sum in Eq. (\ref{Pexpand}). For the fixed permutation, we can express
the product $\Psi_{220}$ as a polynomial in
$\{ (z_i^\uparrow - z_{\sigma(i)}^\downarrow), (z_i^\uparrow + z_{\sigma(i)}^\downarrow); i = 1, \ldots, N/2 \}$
variables. In this way, when we combine $\Psi_{220}$ with the pairing part with fixed $\sigma$,
the projection reduces to the projection of each $\frac{z^m}{\bar z}$ (with $m = 0, \ldots, N - 2$), to  the LLL.
If we choose the symmetric gauge, the basis of the LLL is given by  $\Psi_k (z) = \frac{z^k}{\sqrt{2 \pi k! 2^k}}$, with $k = 0, \ldots, N-1$.
Here we dropped the Gaussian factor, $ \exp\{- |z|^2/4\}$, which is included in the measure defining the scalar product \cite{jach}. Thus,
\begin{equation}
\langle \Psi_k| \frac{z^m}{\bar z} \rangle = \delta_{m, k-1} \frac{2 \pi m! 2^m}{\sqrt{2 \pi k! 2^k}}.
\end{equation}
It follows that
\begin{equation}
\mathcal{P}_{LLL}(\frac{z^m}{\bar z}) = \sum_k \langle \Psi_k| \frac{z^m}{\bar z} \rangle  \Psi_k = \frac{z^{m+1}}{2(m+1)}.
\end{equation}
In particular, for $m = 0$, $\mathcal{P}_{LLL}(\frac{1}{\bar z}) = \frac{z}{2}$, i.e., the leading term
in the short distance approximation reproduces the behavior of the $p$-wave paired composite fermion wave function described by Eq.(\ref{sdispairing}). 

\section{Relation to pseudospin spiral order}\label{app:park}

The first exact diagonalization study on the torus 
that pointed out to the existence of the competing 
ground states with diverging momenta in the
thermodynamic limit was done by Park in Ref. \onlinecite{park}.
Our numerical results are in agreement with
Ref. \onlinecite{park}, in particular for $N=13$ electrons
and bilayer distance $d=1.5\ell_B$ [Fig. \ref{Fig:spectrum1315}(a)].

Ref. \onlinecite{park} pointed out that one energy level (the lowest excitation
at non-zero momentum, which represents four states due to  
inversion symmetry) is ``ripped out" from the
quasi-continuum of high momentum states. The momenta that correspond to these states
are the ones in Eq. (\ref{state1}) in the main text, or
\begin{eqnarray}
(N_{\uparrow},0), \; (N_{\downarrow},0), \; (0,N_{\uparrow}) \; 
{\rm and} \; (0,N_{\downarrow}),
\end{eqnarray}
in units of $2\pi/L$ where $L=\sqrt{2\pi \ell_B^2 N}$.

In order to motivate and explain the emergence of the particular
momenta above, Park discussed the ``spiraling states"
in the Landau gauge,
\begin{equation}
|\Psi_{spiral}(n)\rangle = \prod_{n} \frac{1}{\sqrt{2}}
 (c_{m}^{\downarrow\dagger} + c_{m+n}^{\uparrow\dagger}) |0\rangle. 
\label{Pspiral}
\end{equation}
Here  $c_m^\dagger$ denotes the electron creation operator, with
integer $m = 0,\ldots, N-1$ labeling the linear momentum along the cylinder, i.e.,
$\frac{2 \pi m}{L}$. With respect to the ground state of the $XY$ ferromagnet [$n = 0$ in Eq. (\ref{Pspiral}), or
Eq. (\ref{ferro}) if we consider the fixed number of particles representation],
the acquired momentum is $\mathbf{k} \ell_B = (\frac{2\pi n N_\uparrow}{L},0)$.

In the first quantization, in terms of the usual Landau gauge states
\begin{equation}
 \exp{ \left(\frac{2 \pi i m z_k}{L} \right)} \exp{\left(- \frac{1}{2} y_k^2\right)},
\end{equation}
where $z_k = x_k + i y_k$, $k = 1, \ldots,N $,  we can express the
spiraling state as
\begin{equation}
\prod_i \left(
  \begin{array}{c}
    \exp\left( i \frac{2\pi}{L} x_i\right) \\
    1
  \end{array}
\right)
\prod_{i<j} (Z_i - Z_j) \exp\{- \frac{1}{2} \sum_i y_i^2\},
\end{equation}
where we introduced $Z_j = \exp\left(\frac{2 \pi i  z_j}{L}\right)$. Thus the physics of the neutral sector is essentially given by the factor
\begin{equation}
\prod_i \left(
  \begin{array}{c}
    \exp\left(i \frac{2\pi}{L} x_i\right) \\
    1
  \end{array}
\right),
\end{equation}
which describes long spiraling with period $L$ in the pseudospin space. 

Park argued that by more elaborate inclusions of the disordering effects
in the ground state, similar constructions to the ones in (\ref{Pspiral}) 
should be relevant for the description of the new
ground state(s). They will correspond to the new intermediate phase with the
``pseudospin spiral order".

In this work we point out that there are extra states, as seen in Fig. \ref{Fig:spectrum1315}
falling down, which also belong to the set of candidate ground states of the intermediate phase. Given the unusually
large momenta (and long periods) for spiral order, we argue in Sec. \ref{sec:boson}
that they are very likely an expression of the {\em topological}
ordering in the intermediate phase. Furthermore, we connect them 
with merons -- the spin-texture fractional excitations of the $XY$ ordered 
phase -- which are confined (bound in pairs). The long spiraling in the ground states of the intermediate phase 
on the torus is thus naturally explained by the spin-texture description of merons.

\section{\label{Slowlying} Description of the low-lying modes}

To find and understand the description of the low-lying modes it is convenient to begin with the case $d=0$ ($SU(2)$ symmetric case) in the presence of only $V_0$ pseudo potential for which we can identify the whole family of zero modes (zero energy states). As shown in Ref.~\onlinecite{sky} the basic states (from which we can get all other zero-energy states) similarly to edge states in the Laughlin case, can be described in the disc geometry as
\begin{equation}\label{basicstates}
\left(
\sum_{\langle i_1\dots i_L \rangle}
z_{i_1}^\uparrow \dots z_{i_L}^\uparrow
\right)
\prod_{i<j}(z_i-z_j),
\end{equation}
where the sum is over various $L-$tuples of $\uparrow$ particles.

With the inclusion of higher angular momentum pseudo potentials they will not be degenerate anymore, but will constitute the candidate states for the states of the lowest energy at a given momentum $L$ (with respect to the ground state momentum). For example, the state at $L=1$, $(\sum_{i=1}^{N_{\uparrow}}z_{i}^\uparrow)\prod_{i<j}(z_i-z_j)$, can be associated with the first excited state above ground state in the Goldstone branch. This was demonstrated in Ref.~\onlinecite{gmshs} when a boost of angular momentum $L$ acting only on one pseudo-spin projection (layer index) in the ground state was identified with the Goldstone mode with momentum $L.$ This ansatz was very satisfactory for the whole range of distances (from zero to infinity) in the bilayer for low momenta ($L=1,2,3$).

On the other hand the state in Eq.~(\ref{basicstates}) when $L=N_{\uparrow}$
\begin{equation}\label{meronpair}
\prod_{i=1}^{N_{\uparrow}}z_{i}^\uparrow \prod_{i<j}(z_i-z_j),
\end{equation}
is a meron-(anti-meron at the infinity) state. This state is a projection of the meron state introduced in Ref.~\onlinecite{moon}
\begin{equation}
\prod_{n}(c_{n}^{\downarrow\dagger} + c_{n+1}^{\uparrow\dagger}) |0\rangle, \label{cmeron}
\end{equation}
where $c_{n}^{\sigma\dagger}, \sigma = \uparrow, \downarrow$ are electron creation operators in the second-quantized notation,
to the fixed relative number, $N_{\uparrow}-N_{\downarrow},$ of electrons. This state was also used by Park~\cite{park} to model the state(s) that fall(s) from the higher-momentum region when the distance ($d$) increases.

The best of overlap in Ref.~\onlinecite{park} was reached around $d \lesssim \ell_B$ -- after that the overlap was rapidly decreasing. As the author already pointed out, the reason for this is that the used construction is based on the 111 state, i.e. the state in Eq.~\ref{ferro} projected to the fixed number of electrons, and we need to include the changes in the ground state that set in around $d\sim \ell_B.$ Thus we may imitate the approach of Ref.~\onlinecite{gmshs} and model new state and correlations by
\begin{equation}\label{ansatz}
\prod_{i=1}^{N_{\uparrow}}z_{\uparrow i}\Psi_0,
\end{equation}
where $\Psi_0$ represents the exact ground state at distance $d$ that we find in $k=0$ sector.

Nevertheless, the quantum numbers of relevant states, Eq.~\ref{ansatz}, are the same as from the states they evolve from, Eq.~\ref{meronpair}, and we can use these states i.e. their description given by Eq.~\ref{description} to classify all possible low lying states on the torus.

\section{Critical region}

\begin{figure*}[ttt]
  \begin{minipage}[l]{\linewidth}
\includegraphics[width=0.95\linewidth]{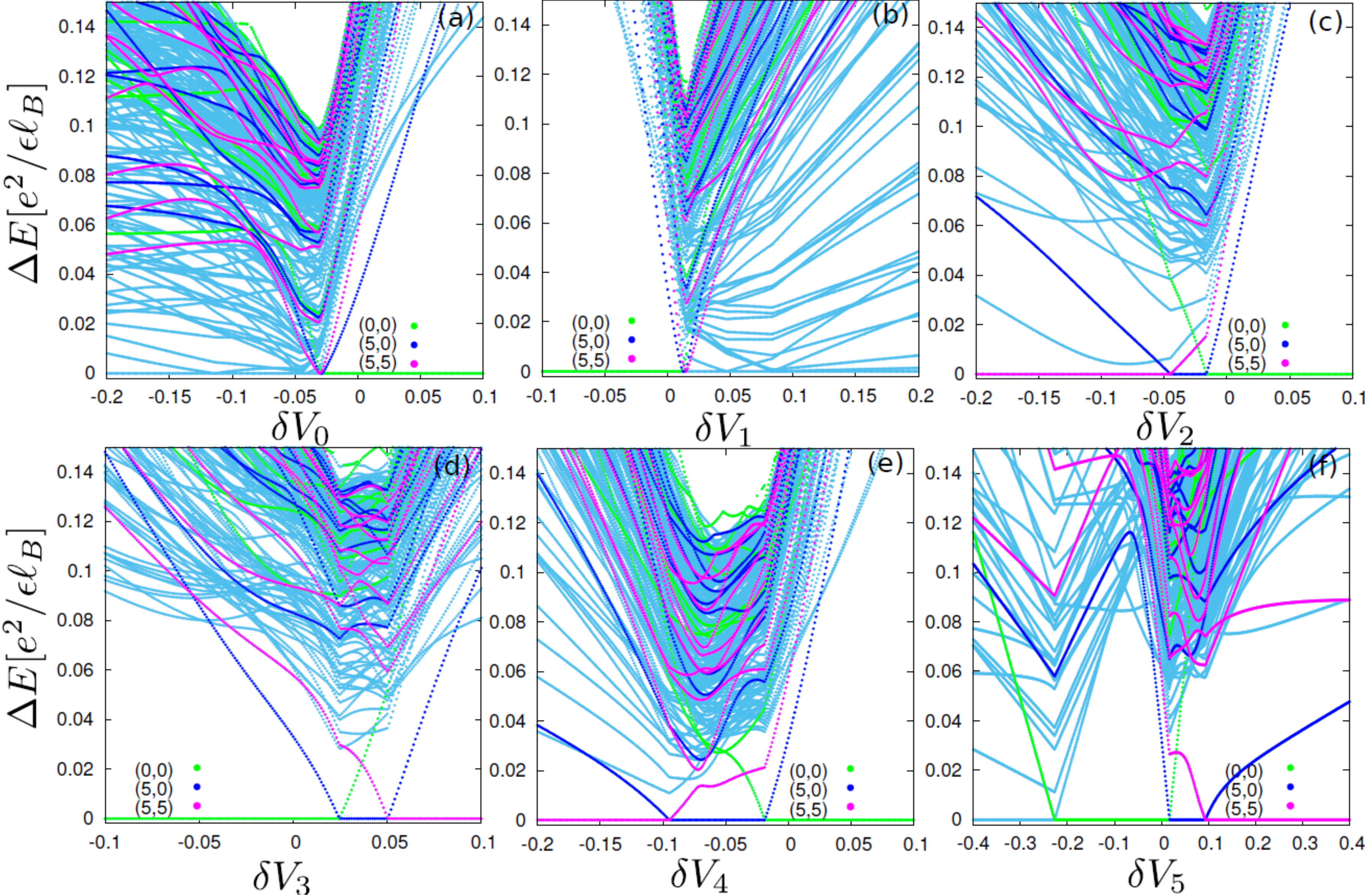}
  \end{minipage}
\caption{(Color Online) Energy spectrum for the Coulomb interaction at bilayer distance $d=1.5\ell_B$ modified by short-range Haldane pseudopotentials $\delta V_m$, $m=0,\ldots 5$ [(a)-(f)]. In all cases, total number of electrons is $N=11$ and the unit cell is a square. The spectrum is obtained by finding 10 lowest eigenvalues in each momentum sector, and for clarity purposes the resulting energies are plotted relative to the ground state for each value of the pseudopotential. Levels belonging to high-symmetry momentum sectors are denoted by special colors. For odd pseudopotentials $\delta V_1, \delta V_3, \delta V_5$, we modify the intra-layer and inter-layer Coulomb by equal amounts. The even pseudopotentials $\delta V_0, \delta V_2, \delta V_4$ only affect the inter-layer Coulomb because of Fermi statistics.}
\label{Fig:11pseudo}
\end{figure*}

Our analysis suggests that the bilayer intermediate state appears as a critical region
between a ferromagnetically ordered state and a (completely) disordered state. It is tempting to
expect that by modifying the parameters in the critical region we can reach the $\mathbb{Z}_2 D $ phase (the deconfined phase of $\mathbb{Z}_2$ gauge theory, mentioned in Sec.~\ref{sec:deconf}). This is one of the simplest topological phases that would be realized by strong pairing of composite bosons. Also, it is important to understand better the influence of the various Haldane pseudopotentials on the competing ground states in the critical region.

In order to explore the phase diagram more broadly, we consider modifications of the bare Coulomb potential by the short-range Haldane pseudopotential terms \cite{prange}:
\begin{equation}
V \to V + 2(2\pi)\sum_m \delta V_m L_m(q^2),
\end{equation}
where $m=0,1,2,\ldots$. The purpose of these modifications is to model the perturbations that occur in an actual experimental system (for example, effects due to finite width of the sample in the perpendicular direction, mixing with higher Landau levels, etc.). Note that we are free to separately tune the intra- and inter-layer interactions. Adding even $m$ pseudopotentials will only have an effect on inter-layer potential because the Fermi statistics inside each layer precludes even values of $m$.

In Fig. \ref{Fig:11pseudo} we can see the evolution of the spectrum with small departures $ \{ \delta V_m ; m = 0, \ldots, 5 \}$ of the Haldane pseudopotentials from their Coulomb values. The influence of the even pseudopotentials, $\delta V_0, \delta V_2,$ and $ \delta V_4$ is qualitatively the same; for stronger positive values the ferromagnetically ordered state is stabilized, and for negative values we expect a compressible behavior. The influence of $ \delta V_1$ is according to our expectations; for stronger positive $ \delta V_1$ we see compressible behavior, while for negative $ \delta V_1$  the ordered state is stabilized.

On the other hand the influence of positive $ \delta V_3$  and $ \delta V_5$ is interesting as it stabilizes one of the set of states (competing ground states) that have characteristic momenta that are compatible with $\mathbb{Z}_2 D $. Nevertheless, a closer inspection of the influence of $ \delta V_3$ does not lead to a firm (topological) characterization of the ground states for positive  values of $ \delta V_3$. Moreover, the behavior of the gap in this region is strongly dependent on system size, which further suggests the phase is compressible.

\acknowledgments

We acknowledge useful discussions with Kwon Park. This work made use of the facilities of N8 HPC provided and funded by the N8 consortium and EPSRC (Grant No.EP/K000225/1). The work was supported by the Ministry of Education, Science, and Technological Development of the Republic of Serbia under projects ON171017 and ON171027.


\begin{thebibliography}{99}
\bibitem{comment} A possible reason for the presence of composite fermions can be seen in a formal way by also noting that, by considering charge and pseudospin sector separately, we have to allow quasiparticle constructions which are not single-valued functions of electron coordinates (assuming that the gluing of
charge and pseudospin will heal this). Thus in the charge sector we can have a Laughlin quasihole, $ \prod (z_i^\uparrow - w) \prod (z_j^\downarrow - w) \Psi_{gs}$, but also $ \sqrt{\prod (z_i^\uparrow - w) \prod (z_j^\downarrow - w)} \Psi_{gs}$, and in the pseudospin sector, ``half", $ [\sqrt{\prod (z_i^\uparrow - w)}/\sqrt{\prod (z_j^\downarrow - w)}] \Psi_{gs}$, and thus also $ [\prod (z_i^\uparrow - w)/\prod (z_j^\downarrow - w)] \Psi_{gs}$
(and other fused quasiparticles, see Appendix \ref{sec:appA}). If we allow for all (single-valued) combinations of these basic quasiparticle constructions we may expect
$ \prod (z_i^\uparrow - w)^2 \Psi_{gs}$ (composite fermion), etc.
\end{thebibliography}
\end{document}